\documentclass[%
 reprint,
 amsmath,amssymb,
 aps,
]{revtex4-1}

\usepackage{graphicx}
\usepackage{dcolumn}
\usepackage{bm}
\usepackage{float} 

\newcommand*{\affaddr}[1]{#1} 
\newcommand*{\affmark}[1][*]{\textsuperscript{#1}}

\begin{document}

\preprint{APS/123-QED}

\title{Connecting Particle Physics and Cosmology: Measuring the Dark Matter Relic Density in Compressed Supersymmetry at the LHC} 

\author{
Carlos Avila\affmark[3], Andr\'es Fl\'orez\affmark[3], Alfredo Gurrola\affmark[1], Dale Julson\affmark[1,2], Savanna Starko\affmark[1]\\
\affaddr{\affmark[1] Department of Physics and Astronomy, Vanderbilt University, Nashville, TN, 37235, USA}\\
\affaddr{\affmark[2] Department of Physics, Texas State University, San Marcos, TX, 78666, USA}\\
\affaddr{\affmark[3] Physics Department, Universidad de los Andes, Bogot\'a, Colombia}\\
}

\date{\today}

\begin{abstract}
The identity of Dark Matter (DM) is one of the most captivating topics in particle physics today. 
The R-parity conserving Minimal Supersymmetric Standard Model (MSSM), which naturally provides a DM candidate 
in the form of the lightest neutralino ($\tilde{\chi}_{1}^{0}$), is used as a benchmark scenario to show that a 
measurement of $\Omega_{\tilde{\chi}_{1}^{0}}h^{2}$ can be achieved from measurements at the CERN Large Hadron Collider. 
Focus is placed on compressed mass spectra regions, where the mass difference between the $\tilde{\chi}_{1}^{0}$ and 
the $\tilde{\tau}_{1}$ is small and where the $\tilde{\tau}_{1}$-$\tilde{\chi}_{1}^{0}$ coannihilation (CA) mechanism of the early 
Universe plays an important role. The technique for measuring $\Omega_{\tilde{\chi}_{1}^{0}}h^{2}$ relies on two 
proposed searches for compressed Supersymmetry (SUSY): 1) production via Vector Boson Fusion (VBF) processes; and 2) 
production with associated energetic jets from initial state radiation (ISR). These approaches allow for the determination of the 
relic abundance at the LHC for any model where CA is an important DM reduction mechanism in the early Universe. Thus, 
it is possible to confirm that the DM we observe today were $\tilde{\chi}_{1}^{0}$'s created in the early Universe. We show 
that from measurements in the VBF and ISR SUSY searches at the LHC, the dark matter relic density can be measured 
with an uncertainty of 25\% with 3000 fb$^{-1}$ of 13 TeV proton-proton data.
\end{abstract}

\pacs{Valid PACS appear here}
\maketitle

One of the most incredible discoveries in modern cosmology is that the Observable Universe is only a small fraction of the total energy density of the Universe~\cite{spergelASTRO}. 
Only 5\% of the Universe's energy density is so-called ``normal matter'', 27\% is Dark Matter (DM), and 68\% is Dark Energy~\cite{planckASTRO}. 
Given the depth of our scientific understanding of the surrounding world, this is a surprise. While there is evidence for DM revealed through its gravitational interactions at macroscopic scales~\cite{DMevidence}, the unknown identity of DM 
is one of the most profound questions in science.
If we can understand the nature of DM, we would make great progress in understanding the evolution of the Universe and its composition.\\
\indent Becoming increasingly clear is the significant overlap in the sciences of the biggest and smallest things in the Universe. Pertinent scientific disciplines - particle and nuclear physics, astronomy, and cosmology - often use similar theories and tools to develop an understanding of the Universe. 
While astronomers seek the effects of DM in space, particle physicists are trying to develop experimental techniques to produce and discover the DM particle at colliders. 
The ultimate goal is to discern what this new form of matter is, measure its properties, and determine how measurements at colliders can be used to deduce today's DM content necessary to understand the evolution of the Universe.\\
\indent Hadron particle colliders have proven to be successful experimental tools to understand the smallest building blocks and fundamental forces. As proof, 
the ATLAS~\cite{Aad:2008zzm} and CMS~\cite{Chatrchyan:2008aa} experiments at the CERN Large Hadron Collider (LHC) discovered the Higgs boson~\cite{Aad20121,Chatrchyan201230}, which is responsible for generating the mass of particles that make up matter. Through these successes, the search for the particle identity of DM has proven to be difficult due to its weakly interacting nature and potentially large mass. 
Its weakly interacting nature means DM is produced in hadron colliders at a significantly lower rate than other known processes mediated by the strong force and producing a similar detector signature. 
The potentially large mass of the DM particle also suppresses its production rate. Particle physicists have realized that potentially the only ways to detect DM at a hadron collider are to target such rare production mechanisms that, although giving low production rates, result in a signature so distinct that DM could be identified amongst more abundantly produced processes.\\
\indent The tagging of events using a Vector Boson Fusion (VBF) topology \cite{VBF1,VBF2,DMmodels2,Khachatryan:2016mbu,VBFSlepton,VBFStop,VBFSbottom} or a highly energetic jet from initial state ration (ISR) \cite{isrstauPHYS}, have been proposed as two of those rare experimental handles to discover the DM particle at the LHC (and other new physics~\cite{Andres:2016xbe,VBFhn2017}).
The use of VBF tagging is effective at suppressing SM backgrounds from QCD processes that dominate at a hadron collider. The use of a high $p_{T}$ ISR jet boosts the detection of DM particles due to their recoiling effect against the ISR jets. However, even if particle $X$ is discovered at the LHC with the proposed VBF or ISR analyses, it is not sufficient to claim $X$ as the DM particle until its relic density is shown to be consistent with the one measured by astronomers. On the other hand, if $X$ is discovered with the VBF/ISR analyses and the deduced relic density is not consistent with the measurements from astronomy, this does not necessarily mean $X$ is not the DM particle. Instead, it could mean our assumptions of the evolution of the Universe (Big Bang Cosmology) might not be correct (e.g. thermal vs non-thermal cosmology). In either case, a discovery at the LHC with the VBF/ISR DM analyses \textit{and the subsequent determination of its relic density} has the potential to 
paint a more comprehensive picture of the DM particle interactions which existed in the early Universe and led to its current structure.\\
\indent The current belief is that at the inception of the Universe, DM particles could be created and destroyed concurrently. According to standard Big Bang Cosmology, temperatures were high enough that Standard Model (SM) particles had enough thermal kinetic energy to interact, annihilate, and produce DM~\cite{brandenbergerASTRO}. Additionally, prior to the Universe's inflation, the concentration of DM was high enough that the DM particle content could be reduced as DM particles underwent interactions producing SM particles. As time progressed, the Universe expanded in the inflationary period, cooling temperatures in the process. The rate of DM creation diminished to practically zero as SM particles lost kinetic energy. In the expanded Universe, DM concentration became more diffuse, diminishing the rate of DM reduction. Since that critical time, the DM density has remained relatively constant~\cite{brandenbergerASTRO}. This measure is referred to as the DM relic density, which is quoted as $\Omega_{\mathrm{DM}}h^{2} \approx 0.11$~\cite{planckASTRO}. However, in some particle physics models, particularly Supersymmetry (SUSY)~\cite{SUSY1,SUSY2,SUSY3,PranNath,SUSY4,SUSY5}, $SM+SM \leftrightarrow DM+DM$ interactions are not sufficient to give way to $\Omega_{\mathrm{DM}}h^{2} \approx 0.11$ (e.g. result in overabundance of DM). In these cases, a model of coannihilation (CA) is necessary to maintain consistency between particle physics and cosmology~\cite{bakerJHEP}. According to CA theory, the DM particle has a CA partner $\Upsilon$, perhaps yet to be discovered, which can interact with the DM particle in the early Universe and coannihilate to produce SM particles, thus reducing the DM content in a way that is consistent with current cosmological measurements made by the Wilkinson Microwave Anisotropy Probe (WMAP)~\cite{spergelASTRO} and the Planck collaboration~\cite{planckASTRO}. Since the CA cross-section depends exponentially on the relative mass splitting $\Delta m$ between DM and $\Upsilon$, $\left< \sigma v \right>_{\textrm{CA}} \sim e^{-\Delta m}$, the CA mechanism of the early Universe becomes important for DM physics in models with compressed mass spectra~\cite{boltzmannPHYS}. These compressed mass spectra regions are hallmark scenarios for the aforementioned VBF and ISR DM search methods at the LHC.\\
\indent \textit{In this Letter we propose a series of measurements at the LHC, using the VBF and ISR search channels, to determine the role of CA, deduce the masses of relevant particles (e.g. DM mass), and establish a prediction of the DM relic density $\Omega_{\mathrm{DM}}h^{2}$ given current and future luminosities at the LHC}. A strong candidate for the DM particle is the lightest neutralino ($\widetilde{\chi}^{0}_{1}$) in the R-parity conserving Minimal Supersymmetric Standard Model (MSSM). The lightest neutralino may be some linear combination of higgsino, bino, and wino - the SUSY equivalents of the SM gauge bosons~\cite{SUSY5}. In our benchmark scenario we assume the $\widetilde{\chi}^{0}_{1}$ has a large bino component, since this is the SUSY phase space where the CA mechanism plays an important role. 
Focus is placed on regions where the mass difference $\Delta m$ between the stau ($\tilde{\tau}$) and the $\widetilde{\chi}^{0}_{1}$ is small ($\sim 5$-25 GeV)~\cite{Pran2}, and where the SUSY $\tilde{\tau}$ is produced through cascade decays of the lightest chargino $\widetilde{\chi}^{\pm}_{1}$ and the next-to-lightest neutralino $\widetilde{\chi}^{0}_{2}$ in processes such as $\widetilde{\chi}^{+}_{1}\widetilde{\chi}^{-}_{1}\to\tilde{\tau}\tilde{\tau}\nu_{\tau}\nu_{\tau}$, $\widetilde{\chi}^{0}_{2}\widetilde{\chi}^{0}_{2}\to\tilde{\tau}\tilde{\tau}\tau\tau$, $\widetilde{\chi}^{\pm}_{1}\widetilde{\chi}^{0}_{2}\to\tilde{\tau}\nu_{\tau}\tilde{\tau}\tau$ and $\widetilde{\chi}^{\pm}_{1}\widetilde{\chi}^{0}_{1}\to\tilde{\tau}\nu_{\tau}\widetilde{\chi}^{0}_{1}$. The colored SUSY sector is decoupled. 
This choice of SUSY mass spectra is driven by two factors: 1) compressed mass spectra regions are difficult to probe at the LHC due to experimental constraints for events containing low $p_{T}$ objects; 
2) \textit{a technique for the precision measurement of $\Omega_{\widetilde{\chi}^{0}_{1}}h^{2}$ at the LHC in this $\tilde{\tau}$-$\widetilde{\chi}^{0}_{1}$ CA region with a decoupled colored sector has not yet been developed}. 
In general, determining the mass and composition of the $\widetilde{\chi}^{0}_{1}$ and also measuring $\Delta m$ in colored cascade searches requires reconstructing several kinematic endpoints simultaneously and in most cases requires one to assume model dependent correlations (e.g. using grand unification to link the mass of the colored sector to the electroweak sector)~\cite{arnowittPHYS}. Furthermore, the ATLAS and CMS experiments have now pushed the limits of the $1^{\textrm{st}}$ and $2^{\textrm{nd}}$ ($3^{\textrm{rd}}$) generation squarks and gluinos to $\sim$ 2.1 TeV (1 TeV)~\cite{SusyColoredSearch,CMS13TeVSusyColoredAllHadSearch2016data}, making the $\widetilde{\chi}^{0}_{1}$ less accessible using colored SUSY searches. Therefore, in this Letter we are motivated by the need for a less model-dependent methodology to deduce $\Omega_{\widetilde{\chi}^{0}_{1}}h^{2}$ in the difficult to probe compressed electroweak SUSY phase space where the CA mechanism of the early Universe is important.\\
\indent The targeted compressed mass spectra scenario results in final states with multiple $\tau$ leptons with low-$p_{T}$ visible decay products ($p_{T} \sim \Delta m$), which makes it difficult to reconstruct and identify more than one. We utilize two search channels which have been proposed as effective probes of the $\tilde{\tau}$-$\widetilde{\chi}^{0}_{1}$ CA region: 1) the invisible VBF channel with events where the visible $\tau$ decay products are too soft to reconstruct and identify but where the two high-$p_{T}$ forward jets boost the missing transverse energy ($E_{T}^{miss}$) to allow for an experimental trigger with low rate and sufficient SM background rejection~\cite{DMmodels2,Khachatryan:2016mbu}; 2) events with one high-$p_{T}$ ISR jet and exactly one $\tau_{h}$ (hadronic decay of the $\tau$ lepton), where the ISR jet boosts the system such that the transverse momentum of the $\tau_{h}$ is large enough to reconstruct and identify experimentally~\cite{isrstauPHYS}.\\
\indent The signal and background samples are generated with MadGraph (v2.2.3) \cite{MADGRAPH}, interfaced with PYTHIA (v6.416) \cite{Sjostrand:2006za} to include quark and gluon fragmentation processes, and Delphes (v3.3.2) \cite{deFavereau:2013fsa} to account for detector effects. 
The first of two sets of signal samples considers the pair production of electroweak SUSY particles 
with an associated ISR jet ($\widetilde{\chi}_{l}^{\pm,0}\widetilde{\chi}_{k}^{\pm,0} j$). The second set contains the same pairs of electroweak SUSY particles, except the production proceeds through the fusion of two SM vector bosons ($W^{\pm}, Z^{0}, \gamma$) and results in two associated forward jets ($\widetilde{\chi}_{l}^{\pm,0}\widetilde{\chi}_{k}^{\pm,0} j_{f}j_{f}$). The signal scans consider $\widetilde{\chi}_{1}^{\pm}$ masses ranging from 100 to 400 GeV, $\Delta m$ values between 5 and 25 GeV, and $\widetilde{\chi}_{1}^{0}$ masses as low as 100 GeV. We select a benchmark reference point where $m(\tilde{\chi}_{1}^{\pm})_{\textrm{benchmark}} = 200$ GeV, $m(\tilde{\chi}_{1}^{0})_{\textrm{benchmark}} = 150$ GeV, and $\Delta m_{\textrm{benchmark}} = 15$ GeV. 
Background events are generated for the production of $W$, $Z$, top-quark pairs ($t\bar{t}$), and vector boson pairs (diboson) with up to four associated jets. Jet matching is performed with the MLM algorithm \cite{MLM}, which requires the optimization of xqcut and qcut. The xqcut defines the minimal distance required among partons at generation level, while the qcut represents the minimum energy spread for a clustered jet in PYTHIA. The optimization of these two parameters is performed using the differential jet rate distribution from MADGRAPH, requiring a smooth transition between the curves for events with $n-1$ and $n$ jets. The optimized xqcut is 15. For the $W$+jets and $t\bar{t}$+jets backgrounds, a qcut of 35 GeV was obtained, while 30 GeV is used for the $Z$+jets background. Finally, a minimal event selection criteria is applied at generation level, which requires leptons to have $p_{T} (\ell) > 10$ GeV and $|\eta (\ell)| < 2.5$, while jets have a minimum $p_{T}$ threshold of 20 GeV and $|\eta| < 5.0$.\\
\indent For the VBF invisible search region, we follow suggested cuts in references~\cite{DMmodels2,Khachatryan:2016mbu} and select simulated events with two forward highly-energetic jets ($p_{T} > 50$ GeV) in opposite hemispheres of the detector ($\eta_{j_{f,1}}\times\eta_{j_{f,2}} < 0$ and $|\Delta\eta(j_{f,1}j_{f,2})| > 4.2$), and reconstructed dijet mass greater than 750 GeV. Furthermore, events which contain a b-tagged jet with $p_{T} > 20$ GeV or an isolated light lepton with $p_{T} > 10$ GeV ($> 15$ GeV for $\tau_{h}$) and $|\eta| < 2.5$ are rejected.\\
\indent The selection criteria used for the ISR $+$ $1\tau_{h}$ channel are similar to those of reference~\cite{isrstauPHYS}. Simulated events are required to have one $\tau_{h}$ with $|\eta| < 2.3$ and $p_{T}$ between 15 and 35 GeV. We veto events containing a b-tagged jet or an isolated light lepton. The leading jet for events which accumulate the search region must have $p_{T} > 100$ GeV and $|\eta| < 2.5$. To ensure the ISR $+$ $1\tau_{h}$ channel is exclusive to the VBF search sample, we veto events which contain a second jet with $p_{T} > 50$ GeV and which can be combined with the leading jet to satisfy the VBF cuts. 
As a cross-check, the results from references~\cite{DMmodels2,Khachatryan:2016mbu,isrstauPHYS} have been reproduced at a level of agreement of $< 20$\%.\\
  \begin{figure}[]
 \begin{center} 
 \includegraphics[width=0.45\textwidth, height=0.37\textheight]{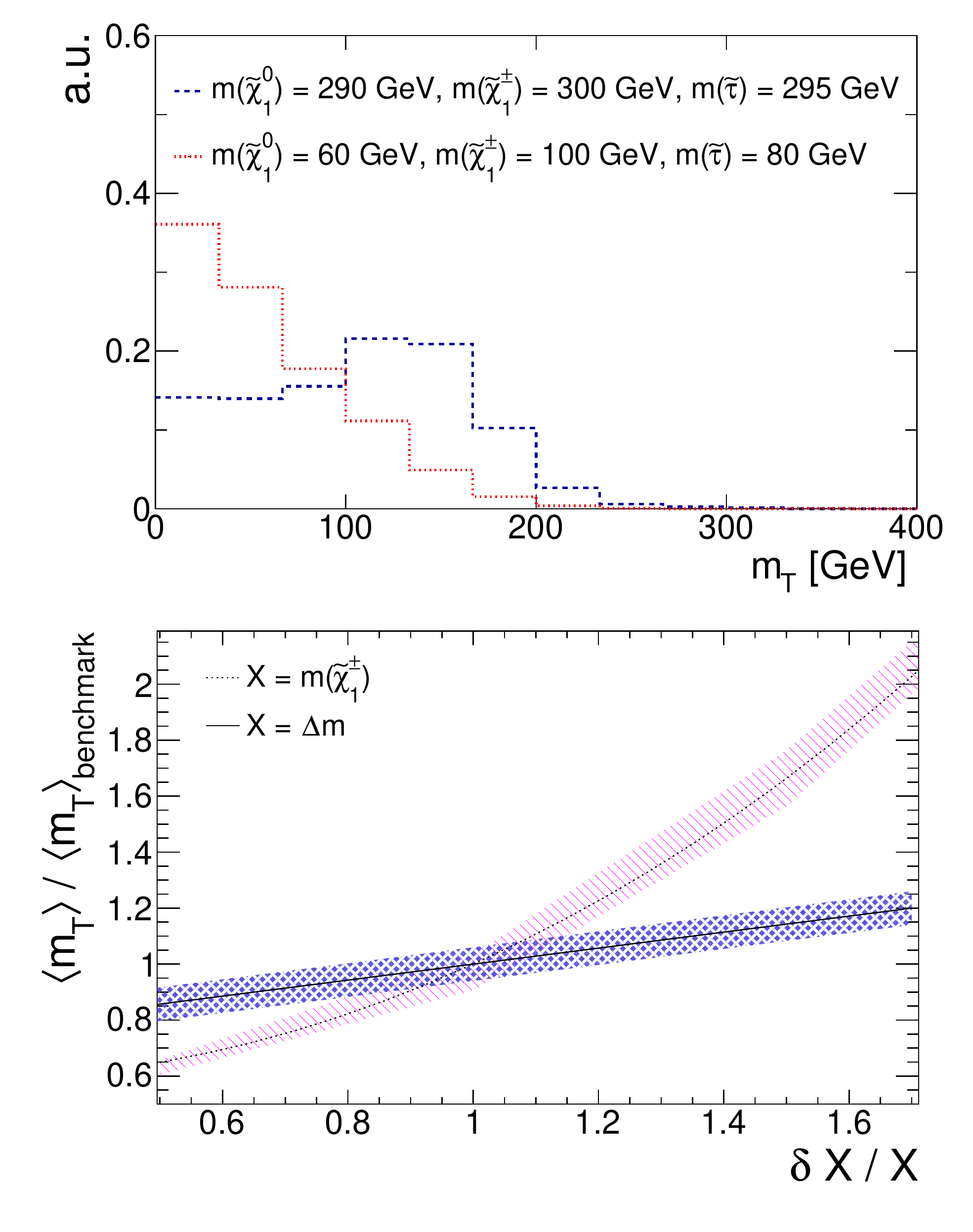}
 \end{center}
 \caption{The top panel shows the $m_{T}$ distributions (normalized to unity) for two different $\tilde{\chi}_{1}^{\pm}$ masses, while the lower panel displays $\left< m_{T} \right>$ as a function of $m(\tilde{\chi}_{1}^{\pm})$ and $\Delta m$, keeping $m(\tilde{\chi}_{1}^{0})$ constant.}
 \label{fig:mTmeanVsChiM}
 \end{figure} 
 \begin{figure}[]
 \begin{center} 
 \includegraphics[width=0.45\textwidth, height=0.37\textheight]{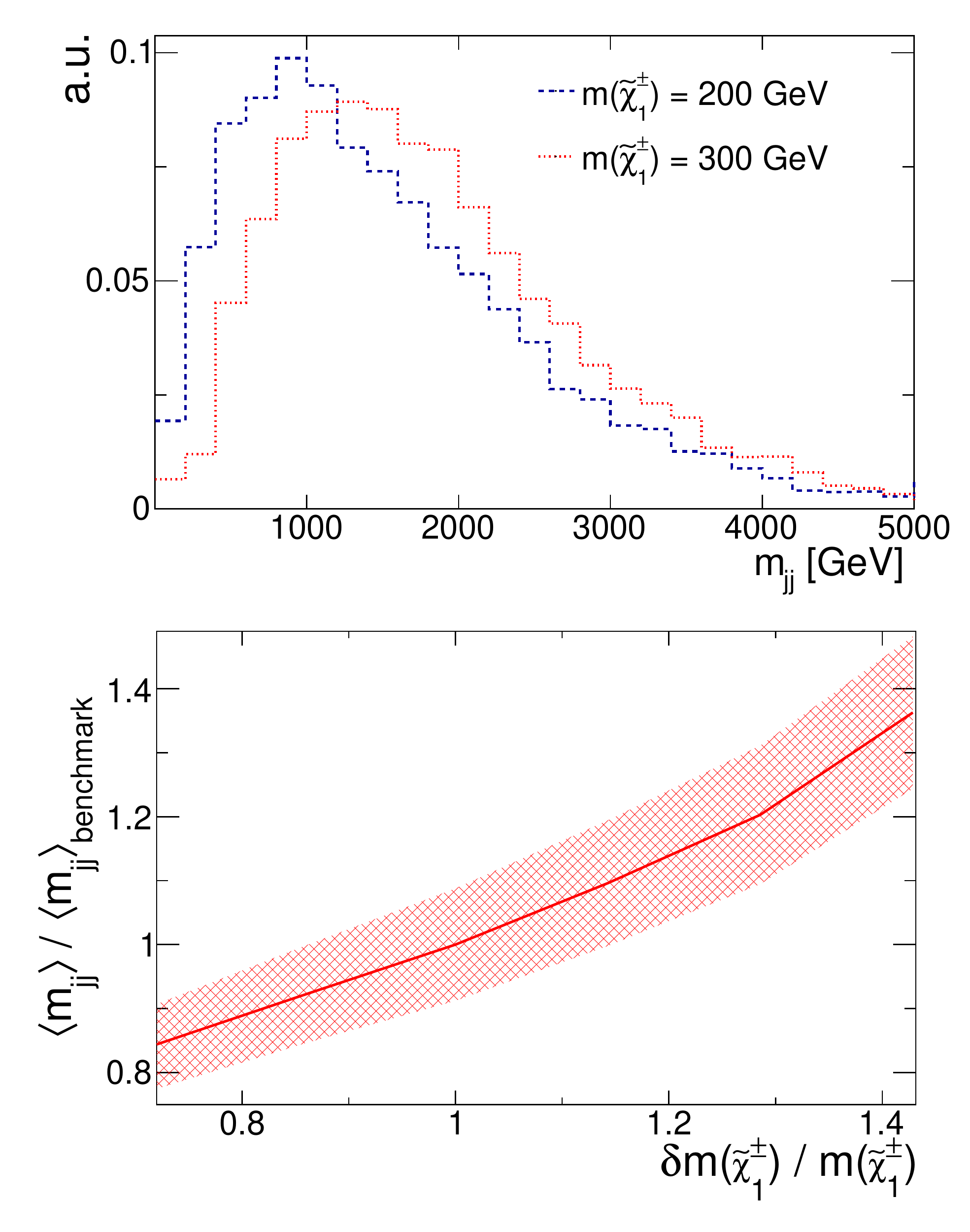}
 \end{center}
 \caption{The top panel shows the $m_{jj}$ distributions (normalized to unity) for two different $\tilde{\chi}_{1}^{\pm}$ masses, while the lower panel displays $\left< m_{jj} \right>$ as a function of $m(\tilde{\chi}_{1}^{\pm})$, keeping $\Delta m$ and $m(\tilde{\chi}_{1}^{0})$ constant.}
 \label{fig:mJJmeanVsChiM}
 \end{figure} 
\indent The VBF invisible search channel uses the reconstructed mass between the two forward jet candidates, defined in Equation \ref{eq:mjjReco}, as the discriminating variable to look for an enhancement of events in the tails of the distribution that could suggest new physics. The ISR $+$ $1\tau_{h}$ analysis uses the reconstructed transverse mass, defined in Equation \ref{eq:mTReco}, between the $\tau_{h}$ and the $E_{T}^{miss}$ from the undetected $\widetilde{\chi}_{1}^{0}$s.
\begin{equation}
  m_{j_{f}j_{f}} = \sqrt{2p_{T}(j_{f1})p_{T}(j_{f2}) cosh\Delta \eta(j_{f1}, j_{f2})}
  \label{eq:mjjReco}
\end{equation}
\begin{equation}
  m_{T} = \sqrt{2 E^{miss}_{T} p_{T}(\tau_{h}) ( 1 - cos\Delta \phi(E^{miss}_{T}, \tau_{h}))}
  \label{eq:mTReco}
\end{equation}
Because of momentum and energy conservation, there is a kinematic correlation between $p_{T}(j)$, $E^{miss}_{T}$, and $p_{T}(\tau_{h})$. In signal events, production of heavier electroweak SUSY particles requires jets with higher $p_{T}$. Therefore, the means of the $m_{j_{f}j_{f}}$ and $m_{T}$ signal distributions ($\left< m_{jj} \right>$ and $\left< m_{T} \right>$) depend on $m(\tilde{\chi}_{1}^{\pm})$. Additionally, because $p_{T}(\tau_{h})$ and $E^{miss}_{T}$ depend on the mass difference $\Delta m$ as well as $m(\tilde{\chi}_{1}^{0})$, the mean of the $m_{T}$ distribution also depends on $\Delta m$ and $m(\tilde{\chi}_{1}^{0})$. The top panel of Figure~\ref{fig:mTmeanVsChiM} shows the $m_{T}$ distributions for two different signal points, while the lower panel of Figure~\ref{fig:mTmeanVsChiM} displays $\left< m_{T} \right>$ (normalized by $\left< m_{T} \right>_{benchmark}$) as a function of $m(\tilde{\chi}_{1}^{\pm})$ and $\Delta m$, keeping $m(\tilde{\chi}_{1}^{0})$ constant. Similarly, the top panel of Figure~\ref{fig:mJJmeanVsChiM} shows the $m_{jj}$ distributions for two different $\tilde{\chi}_{1}^{\pm}$ masses, while the lower panel displays $\left< m_{jj} \right>$ (normalized by $\left< m_{jj} \right>_{benchmark}$) as a function of $m(\tilde{\chi}_{1}^{\pm})$, keeping $m(\tilde{\chi}_{1}^{0})$ and $\Delta m$ constant. The bands in Figures~\ref{fig:mTmeanVsChiM}-\ref{fig:mJJmeanVsChiM} correspond to the statistical uncertainty on the means, calculated as the RMS of the distribution of interest divided by the poisson error on the signal yield assuming an integrated luminosity of $L_{int} = 50$ fb$^{-1}$. 
The means $\left< m_{jj} \right>$ and $\left< m_{T} \right>$ are parameterized as functions of the relevant masses: $\left< m_{jj} \right> = f_{VBF}(m(\tilde{\chi}_{1}^{\pm}),m(\tilde{\chi}_{1}^{0}),\Delta m)$ and $\left< m_{T} \right> = f_{ISR}(m(\tilde{\chi}_{1}^{\pm}),m(\tilde{\chi}_{1}^{0}),\Delta m)$.
\\ \indent Similarly, the observed number of signal events in each channel is a function of the $\tilde{\chi}_{l}^{\pm,0}\tilde{\chi}_{k}^{\pm,0} j_{f}j_{f}$ and $\tilde{\chi}_{l}^{\pm,0}\tilde{\chi}_{k}^{\pm,0} j$ production cross-sections. Therefore, the signal yields depend on $m(\tilde{\chi}_{1}^{\pm})$ and $m(\tilde{\chi}_{1}^{0})$. The production cross-sections also depend on the ``ino'' composition of the neutralinos and charginos. 
The gaugino mixing is driven by the $\mu$ parameter: decreasing the $\mu$ parameter reduces the $\tilde{\chi}_{1}^{0}$ Bino composition by making the Higgsinos more important, and thus simultaneously decreases the Wino composition of the $\tilde{\chi}_{1}^{\pm}$/$\tilde{\chi}_{2}^{0}$. As noted previously, because the $p_{T}$ spectrum of the $\tau_{h}$ in the ISR search channel depends on the mass difference between the $\tilde{\tau}$ and $\tilde{\chi}_{1}^{0}$, the observed number of signal events also depend on $\Delta m$. Based on the above considerations, the signal yields in the VBF invisible and 
ISR $+$ soft-$\tau_{h}$ channels are parameterized as follows: $N_{VBF} = g_{VBF}(m(\tilde{\chi}_{1}^{\pm}),m(\tilde{\chi}_{1}^{0}),\Delta m, \mu)$ and $N_{ISR} = g_{ISR}(m(\tilde{\chi}_{1}^{\pm}),m(\tilde{\chi}_{1}^{0}),\Delta m, \mu)$.\\
\indent If the CMS and ATLAS experiments observe an excess of events in the VBF invisible and ISR $+$ soft-$\tau_{h}$ channels, the relevant particle masses, gaugino mixing parameter $\mu$, and their uncertainties can be deduced from the ``bumps'' in the $m_{jj}$ and $m_{T}$ distributions. This is accomplished by subtracting the predicted background yields from the data, extracting the means of the resulting $m_{jj}$ and $m_{T}$ ``bumps'', determining the observed signal yields, and then inverting the four functions $\left< m_{jj} \right>{=}f_{VBF}(m(\tilde{\chi}_{1}^{\pm}),m(\tilde{\chi}_{1}^{0}),\Delta m),N_{VBF}{=}g_{VBF}(m(\tilde{\chi}_{1}^{\pm}),$\\$m(\tilde{\chi}_{1}^{0}),\Delta m, \mu),\left< m_{T} \right>{=}f_{ISR}(m(\tilde{\chi}_{1}^{\pm}),m(\tilde{\chi}_{1}^{0}),\Delta m),$ and $N_{ISR}{ = }g_{ISR}(m(\tilde{\chi}_{1}^{\pm}),m(\tilde{\chi}_{1}^{0}),\Delta m, \mu)$. With 500 fb$^{-1}$ of 13 TeV pp data from the LHC, we obtain (in GeV) $m(\tilde{\chi}_{1}^{\pm}) = 200 \pm 6.2$, $m(\tilde{\chi}_{1}^{0}) = 150 \pm 7.8$, $\Delta m = 15.0 \pm 1.7$, and $\mu = 500 \pm 42.0$ for our benchmark scenario. We note that the determination of small $\Delta m$ would confirm that we are indeed in the $\tilde{\tau}$-$\tilde{\chi}_{1}^{0}$ CA region.\\
  \begin{figure}[]
 \begin{center} 
 \includegraphics[width=0.45\textwidth, height=0.3\textheight]{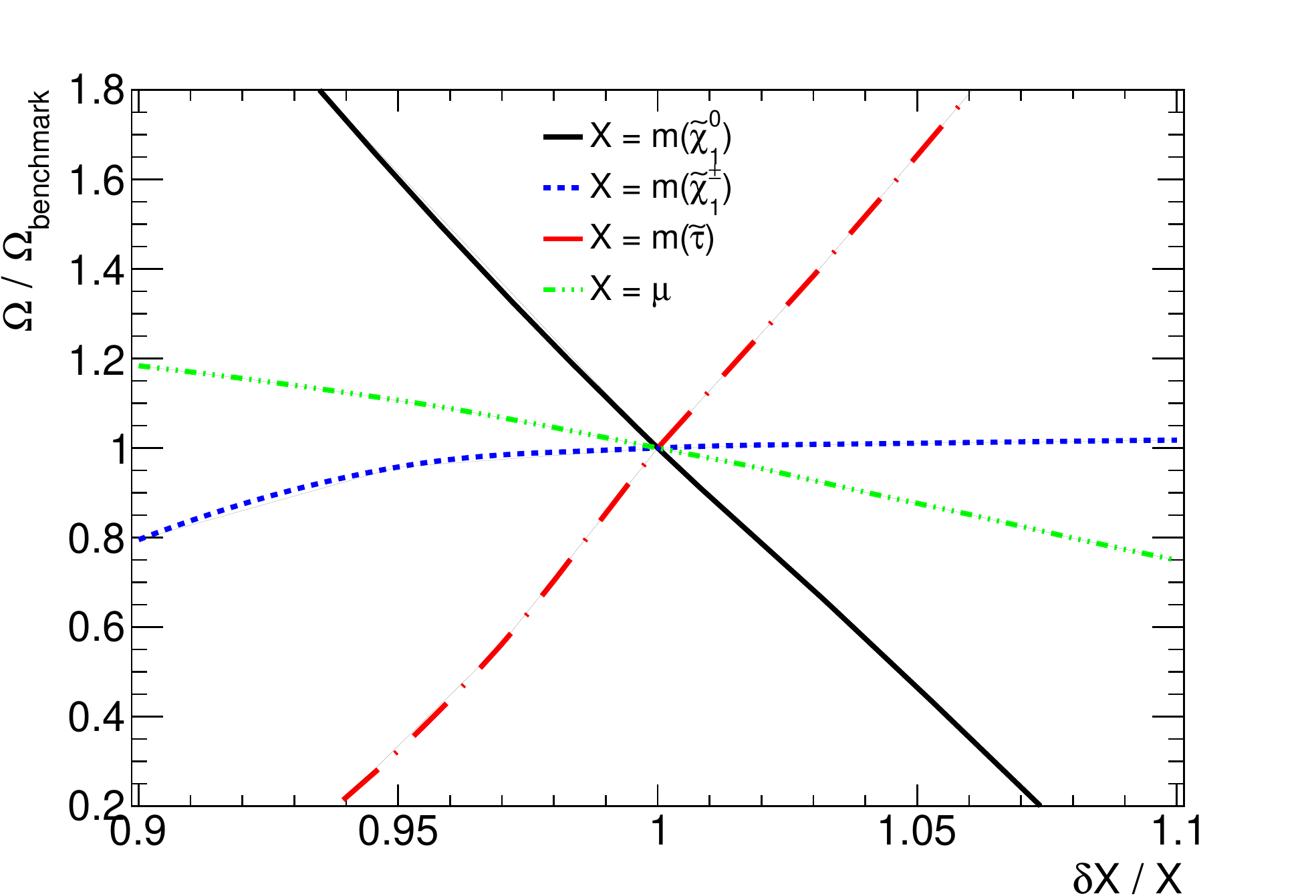}
 \end{center}
 \caption{$\Omega_{\widetilde{\chi}^{0}_{1}}h^{2}$ as a function of $m(\tilde{\chi}_{1}^{\pm})$, $m(\tilde{\chi}_{1}^{0})$, $\Delta m$, and $\mu$. For a given curve, all other parameters are fixed.}
 \label{fig:omegaplot}
 \end{figure}
  \begin{figure}[]
 \begin{center} 
 \includegraphics[width=0.45\textwidth, height=0.37\textheight]{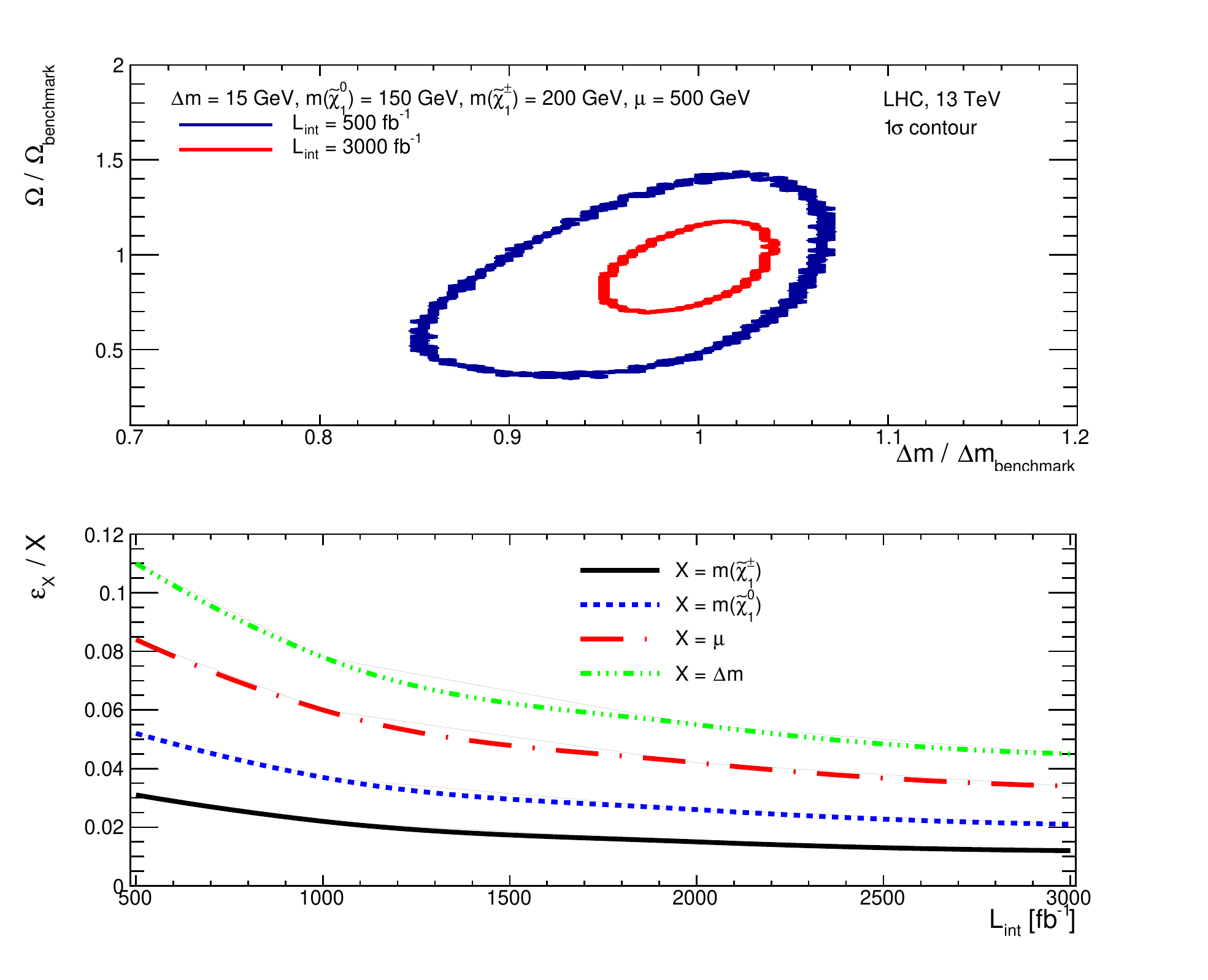}
 \end{center}
 \caption{$\Omega_{\widetilde{\chi}^{0}_{1}}h^{2}$ as a function of $m(\tilde{\chi}_{1}^{\pm})$, $m(\tilde{\chi}_{1}^{0})$, $\Delta m$, and $\mu$. For a given curve, all other parameters are fixed.}
 \label{fig:ellipse}
 \end{figure}
\indent After measuring the sparticle masses and gaugino mixing parameter, we calculate $\Omega_{\widetilde{\chi}^{0}_{1}}h^{2}$ using micrOMEGAs 4.3~\cite{micrOmegas}. The relic density depends on the ``ino'' composition of $\tilde{\chi}_{1}^{0}$, $m(\tilde{\chi}_{1}^{0})$, and $\Delta m$ (since $\left< \sigma v \right>_{\textrm{CA}}$ depends on the Boltzmann factor $e^{-\Delta m}$ in the relic density formula). Figure~\ref{fig:omegaplot} shows $\Omega_{\widetilde{\chi}^{0}_{1}}h^{2}$ as a function of $m(\tilde{\chi}_{1}^{\pm})$, $m(\tilde{\chi}_{1}^{0})$, $\Delta m$, and $\mu$. For a given curve in Figure~\ref{fig:omegaplot}, all other parameters are fixed. For example, the $\Omega_{\widetilde{\chi}^{0}_{1}}h^{2}$ vs. $\Delta m$ curve is obtained by fixing the values $m(\tilde{\chi}_{1}^{\pm})$, $m(\tilde{\chi}_{1}^{0})$, and $\mu$. Similarly, the $\Omega_{\widetilde{\chi}^{0}_{1}}h^{2}$ vs. $\widetilde{\chi}^{0}_{1}$ curve is obtained by fixing the values $m(\tilde{\chi}_{1}^{\pm})$, $\Delta m$, and $\mu$. Since $\Omega_{\widetilde{\chi}^{0}_{1}}h^{2}$ is inversely related to the coannihilation cross-section $\left< \sigma v \right>_{\textrm{CA}} \sim e^{-\Delta m}$, then $\Omega_{\widetilde{\chi}_{0}^{1}}h^{2} \sim e^{\Delta m}$. Thus as $\Delta m$ increases in Figure~\ref{fig:omegaplot}, so does $\Omega_{\widetilde{\chi}^{0}_{1}}h^{2}$. Furthermore, since decreasing the $\mu$ parameter decreases the bino and wino compositions of the $\widetilde{\chi}^{0}_{1}$, the annihilation cross-section $\left< \sigma v \right>$ also decreases, which in turn increases $\Omega_{\widetilde{\chi}^{0}_{1}}h^{2}$. Finally, as $m(\tilde{\chi}_{1}^{0})$ increases, it is less likely that SM particles had enough thermal kinetic energy in the early universe to create heavy DM particles. Therefore, $\Omega_{\widetilde{\chi}^{0}_{1}}h^{2}$ in Figure~\ref{fig:omegaplot} decreases with $m(\tilde{\chi}_{1}^{0})$. \\
\indent Since the DM relic density can be parameterized as $\Omega_{\widetilde{\chi}^{0}_{1}}h^{2} = h(m(\tilde{\chi}_{1}^{\pm}),m(\tilde{\chi}_{1}^{0}),\Delta m, \mu)$, by combining the measurements of the three sparticle mass and gaugino mixing parameter, the DM relic density (and it's uncertainty) can be deduced. Figure~\ref{fig:ellipse} (top panel) shows contour plots of the uncertainty (1 standard deviation) in the $\Omega_{\widetilde{\chi}^{0}_{1}}h^{2}$-$\Delta m$ plane (normalized by the $\Omega_{\widetilde{\chi}^{0}_{1}}h^{2}$ and $\Delta m$ central values for the benchmark signal point). The uncertainty on $\Omega_{\widetilde{\chi}^{0}_{1}}h^{2}$ is 25 (45)\% at 3000 (500) fb$^{-1}$. Figure~\ref{fig:ellipse} (bottom panel) also shows how well the sparticle masses and gaugino mixing can be measured as a function of integrated luminosity. Note the dominant contribution to the uncertainty on $\Omega_{\widetilde{\chi}^{0}_{1}}h^{2}$ is from the measurement of $\Delta m$ (11\% at $L_{int} = 500$ fb$^{-1}$).\\
\indent In conclusion, a technique has been developed for the precision measurement of $\Omega_{\tilde{\chi}_{1}^{0}}h^{2}$ in the $\tilde{\tau}$-$\tilde{\chi}_{1}^{0}$ CA region, using observables from the VBF invisible and ISR $+$ 1$\tau_{h}$ searches for compressed SUSY at the LHC. 
The methodology established in this Letter is agnostic to the mass scale of the colored SUSY sector.
This approach allows for the determination of $\Omega_{\tilde{\chi}_{1}^{0}}h^{2}$ at the LHC for any model where CA is an important DM reduction mechanism in the early Universe. $\Omega_{\tilde{\chi}_{1}^{0}}h^{2}$ can be measured with an uncertainty of 25\% (45\%) with 3000 (500) fb$^{-1}$ of 13 TeV proton-proton data.
This is a critical link between particle physics and cosmology, providing valuable information to help determine whether the DM we observe gravitationally are indeed $\tilde{\chi}_{1}^{0}$s created in the early Universe. 
On the other hand, if $\tilde{\chi}_{1}^{0}$s are discovered with the VBF/ISR analyses and the deduced relic density is not consistent with astronomy, this could lead to a reconsideration of our assumptions of the evolution of the Universe. 
In either case, a discovery at the LHC with the VBF/ISR analyses and the subsequent determination of the $\tilde{\chi}_{1}^{0}$ relic density has the potential to break significant ground on the identity of DM, one of the most relevant questions in science.\\
\indent We thank the constant and enduring financial support received for this project from the faculty of science at Universidad de los Andes (Bogot\'a, Colombia), the administrative department of science, technology and innovation of Colombia (COLCIENCIAS), the Physics \& Astronomy department at Vanderbilt University and the US National Science Foundation. This work is supported in part by NSF Award PHY-1506406.

\end{document}